# Real-time probing of magnetic domain wall dynamic

XING Tao[1,2,4], VERNIER Nicolas[2,3#], ZHANG Xueying[1,4,5*], RASKINE Alessio[3] & ZHAO Weisheng[1]

**AFFILIATIONS**

[1] *Fert Beijing Institute, MIIT Key Laboratory of Spintronics, School of Integrated Circuit Science and Engineering, Beihang University, Beijing 100191, China*

[2] *Centre de Nanosciences et de Nanotechnologies, CNRS, Université Paris-Saclay, 91120 Palaiseau, France*

[3] *Laboratoire Lumière, Matière et Interfaces, CNRS, ENS Paris-Saclay, CentraleSupelec, Université Paris-Saclay, 91190 Gif-sur-Yvette, France*

[4] *Beihang Hangzhou Innovation Institute (Yuhang), Beihang University, Hangzhou 311121, China*

[5] *Truth Instruments Co. Ltd., 266100, Qingdao, China*

[#] nicolas.vernier@u-psud.fr;

[*] xueying.zhang@buaa.edu.cn

## ABSTRACT

We present a study of a very seldom used way of measuring magnetic domain wall velocity, which makes it possible to have a real-time probing of the domain movement in the area illuminated by a laser spot. We have compared this method to the most usual one: although the velocities are similar, the different method do not give the same results if the laser spot is too small. It can be explained by a dendritic shape of the domain wall. By changing the size spot, we propose a basic model which describes quite well the transit time in the laser spot as a function of its size and makes it possible to extract the velocity and the depth of the dendrites.

## I. INTRODUCTION

Propagation of magnetic domain wall is involved in the principle of many potential devices such as the racetrack memory [1-4] or field sensor [5]. This phenomenon is also used in some prototype for logic [6-11]. At last, some people are thinking of using magnetic nanowires to simulate neuronal networks, the nervous influx [12]. So, it is important to understand this phenomenon, and we need efficient method to study it.

Several ways have been used to measure the propagation velocity of a magnetic domain wall in a magnetic thin film. The most common methods consist in, first, creating a domain wall, second, getting a Kerr picture, third, applying a magnetic pulse of known duration and fourth get a second Kerr picture [13]. Then, by measuring the distance of propagation, the velocity can be deduced. This method assumes a negligible inertia of the domain wall, which seems to be all right, as magnetic field pulses are much longer (at least a few µs) than the delays induced by the domain wall inertia (no more than a few ns) [14,15].

Another way of measuring this velocity, is the use of a nanowire with two Hall crosses [16-18]. It requires a nanowire, which means not really 2D propagation. In addition, the Hall crosses can act as pinning positions and one cannot be fully sure of what is obtained. To finish, you get the time reversal at each Hall crosses, but you do not really know what is happening between these two points. For example, you cannot guarantee that a nucleation hasn't occurred in the nanowire, which would mean reversal was not due only to propagation and the deduced propagation velocity is wrong.

In the cases of nanowires, an interesting way to measure propagation velocity is the use of magnetoresistance [18-20]. It assumes the probe current injected is low enough to induce a negligible spin torque in the soft layer [21], which is quite all right as the density of current required to create a meaningful torque is huge [22,23]. However, it requires a nanowire and propagation is not really 2D.

We have studied here another way to probe the propagation velocity of domain walls real-time, which can apply to really 2D films: a polarized laser spot is focused on the sample, and the change of polarization due to Kerr effect is monitored in real-time. This method has already been used a few times, but it was not compared to the usual way of measuring velocity [24,25]. Thanks to our setup which makes it possible to do simultaneously Kerr imaging and real-time probing, we have been able to compare this method to the usual one. The two ways give similar results, but, there is an interesting difference, which seems meaningful of the domain shape.

## II. METHODS

### A. Experimental setup

The setup has been created from a home-made Kerr microscope. This Kerr microscope is a usual one [13]. We have added two mirrors mounted on motorized flip, so that, when these mirrors are in the removed position, the microscope can work as a classical Kerr microscope (see figure 1). The first movable mirror has been put on the light path of the microscope. When in working position, it stops the light microscope and sends a laser beam towards the sample. A lens has been put between the laser and the mirror to get a better focusing of the laser beam on the sample. The second mirror has been put on the detection branch of the microscope, between the beam splitter and the camera. When removed, the usual Kerr picture arrives on the camera detector. When in position, the light is reflected towards a fast light-detecting diode. A lens has been added to focus the light on the active part of the diode. The two movable mirrors can be controlled independently, so that it is possible to have the laser arriving on the sample and the camera still working: we have used this configuration to check the size of the laser spot on the sample.

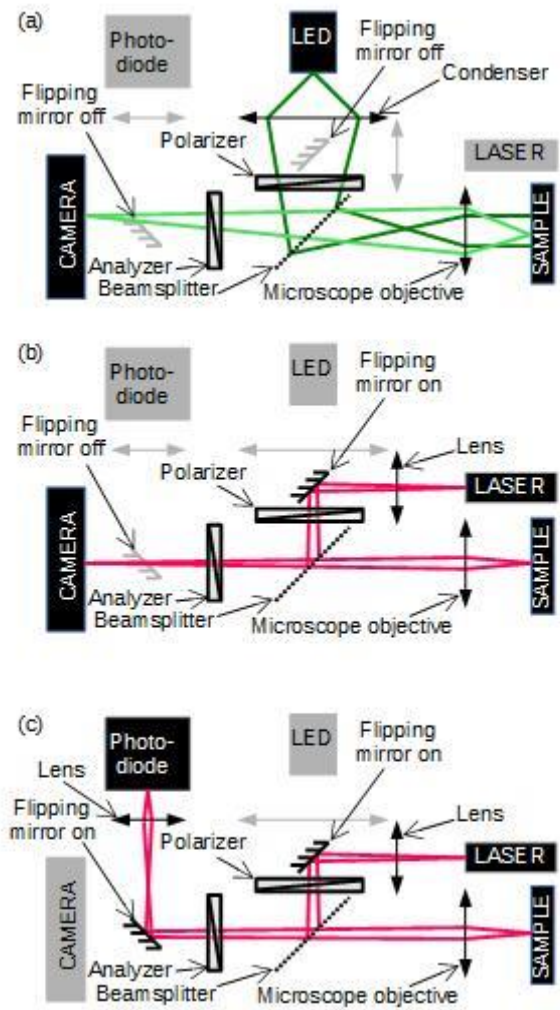

Figure 1. (a) configuration of the setup for Kerr imaging, the two flipping mirrors are removed, the laser is off, the LED is on (b) configuration to check the size of the spot, the first flipping mirror is in position, when the second is still removed, the laser is on , the LED is off, (c) configuration to measure the transit time in the laser spot, both flipping mirrors are in position, the laser is on, the LED is again off.

The signal from the photodiode being very small, a 60dB amplifier has been added after the photodiode. However, because there is a big DC signal, the reversal inducing only a small change of 1%, we have added a RC high pass filter between the photodiode and the amplifier. The cut-off frequency was 1 Hz, which meant negligible effect on the signal due to the filter for a duration shorter than 10 ms. The bandwidth of the amplifier was 100 kHz, limiting the shortest transit times which could be detected to 2.4 µs. An oscilloscope of bandwidth 50 MHz was recording the signal after the amplifier. To make sure not to change the signal because of multi-reflections in the connecting cables, a 50 Ω terminator was put at the entrance of the oscilloscope channel.

The magnetic field pulses were created using a very small coil (not represented in figure 1), put between the sample and the microscope objective. The inductance of the coil was very small, i.e., 150 µH, making it possible to have pulses as short as a few µs (characteristic risetime with 50 Ω adaptation is 1.05 µs). The effective magnetic field could be checked using the second channel of the oscilloscope to monitor the current inside the coil. The sample holder and the coil holder were made of plastic to avoid any eddy current and the magnetic field arising from these eddy current.

The sample was a multilayer Ta(5nm)/Co40Fe40B20(1.1nm)/MgO(1nm)/Ta(5nm) multilayer grown on a Si/SiO$_2$ substrate. It has been annealed at 380°C for 20 minutes. Properties such as hysteresis loop have been already published in a previous paper[20].

## B. Transit time

When applying a magnetic field pulse, we can see very well the reversal on the oscilloscope, as can be seen in figure 2. To get the transit time, we have used the following hypothesis:

- We assumed the laser spot to be Gaussian, with a mean diameter $\omega_0$:

$$I(r) = A.\exp(-2r^2/\omega_0^2) \quad (1)$$

From what we see using the camera, this hypothesis seems good (figure 2)

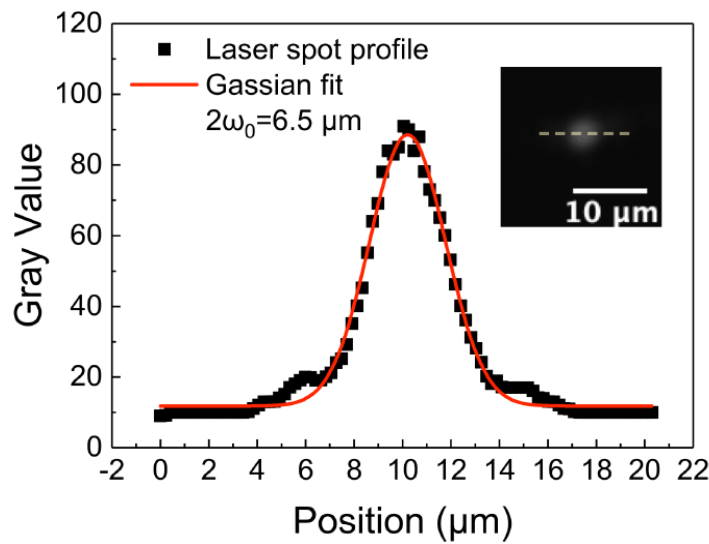

Figure 2. shape of the beam fitted using a Gaussian, the experimental shape is deduced from a profile extracted along the dotted line in the inset, the inset is a picture acquired from the camera using configuration (b) of figure 1

- We assume the domain wall to be a straight line and remain a straight line all through the movement in the spot.
- The velocity is assumed to be constant during the whole transit time in the spot.

These hypothesis leads to a change in the Kerr signal given by [26]:

$$S(t) = K_0 + K_1 \, \text{erf}(\sqrt{2}v(t-t_C)/\omega_0) \qquad (2)$$

This theoretical result is in very good agreement with the experiment, as can be seen in figure 3.

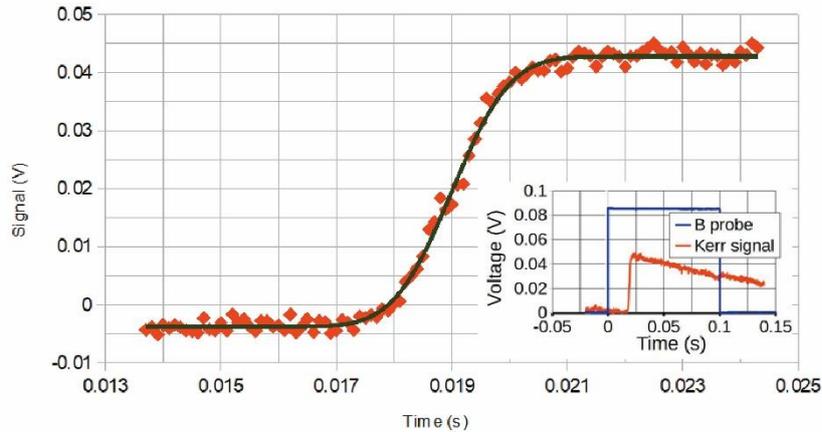

Figure 3. Experimental data recorded with the oscilloscope. The applied field was 1.84 mT. It has been fitted using formula (2). The inset graph shows the whole recorded traces. The blue trace probes the magnetic field, by measuring the voltage on a shunt of 0.2 Ω in serial with the coil. The orange one is the Kerr signal. The decay in the inset curve is due to the high pass filter at the entry of the amplifier. This decay becomes negligible on the small range from 0.014 s to 0.024 s and the effect of the high pass filter can be neglected when fitting the main curve.

Let's note that some software might get troubles to fit the trace using the Erf function. In this case, it is possible to get something satisfying using Tanh instead of Erf, these two functions are very similar. When fitting with $\text{Tanh}(\alpha(t-t_0))$, velocity can be deduced from $v = \alpha\omega_0/1.70$, where r is the radius of the spot defined by the Gaussian beam.

**C. Comparison with velocities obtained through usual way using Kerr pictures**

The first task was to check if the velocity obtained using transit time in the spot gave a result coherent with what we usually get using Kerr pictures. The results have been plotted on figure 4. In the case of the spot measurement, we have used several sizes for the spot.

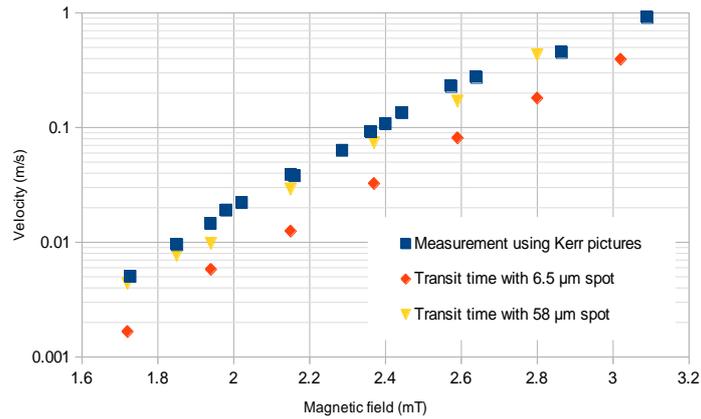

Figure 4. velocities obtained using the usual Kerr pictures way and using our transit time in the spot method. By moving the lens after the laser, it was possible to change the size of the spot, two sizes have been checked in the range [1.8mT; 2.8mT].

On the field range used, which corresponds to the creep regime [27-29], using a logarithmic scale for the vertical one, we can see a shift of the different curves. But, they seem to be parallel. With a spot size of 6.5µm, the shift means a ratio 3 between velocity according to spot measurement and velocity according to Kerr pictures. This ratio is the same on the whole range of field used, although, it seems to become slightly smaller at the highest field. When the size of the spot increases, the spot measurement still gives a velocity lower, but the ratio has reduced to 1.3.

Using the real-time laser spot detection requires to know the size of the spot. Using configuration of figure 1.b, we could get a picture of the laser spot and fit it with a Gaussian law. However, what we get is a picture of the spot through the objective: the size deduced this way might not be the exact one, it might have been enlarged by diffraction. The numerical aperture of the objective was 0.25, it can induce an enlargement of no more than one or two µm. So, with a measured spot size of diameter 6.5 µm, diffraction cannot explain the ratio observed in the experiment.

### III. ANALYSIS

We attribute this difference of velocity to the shape of the domain wall: indeed, to get formula (2), we have assumed the domain wall to be a straight line, but, this hypothesis is wrong. In fact, the domain wall has a dendritic shape, and the successive positions of the domain wall looks like figure 5. When measuring the velocity using Kerr imaging, people measure the velocity propagation of the front part. For example, in figure 5, the bigger spot is reached between t=0 and t=1 the front part and the other side of the spot is reached between t=4 and t=5. As a result, the propagation time required for a distance of 2R, R being the radius of the spot, would be 4 unit time. Using spot detection, the transit time is longer because of the dendritic shape. The signal starts to change between t=0 and t=1 as before, but, the full reversal of the whole area occurs between t=8 and t=9, which means a transit time of 8 unit time instead of 4. As a result, the velocity seems to be two times slower. This ratio should depend on the depth of the dendrites: the longer they are, the bigger the ratio should become, assuming the width of the dendrites to be smaller than the radius of the spot probe.

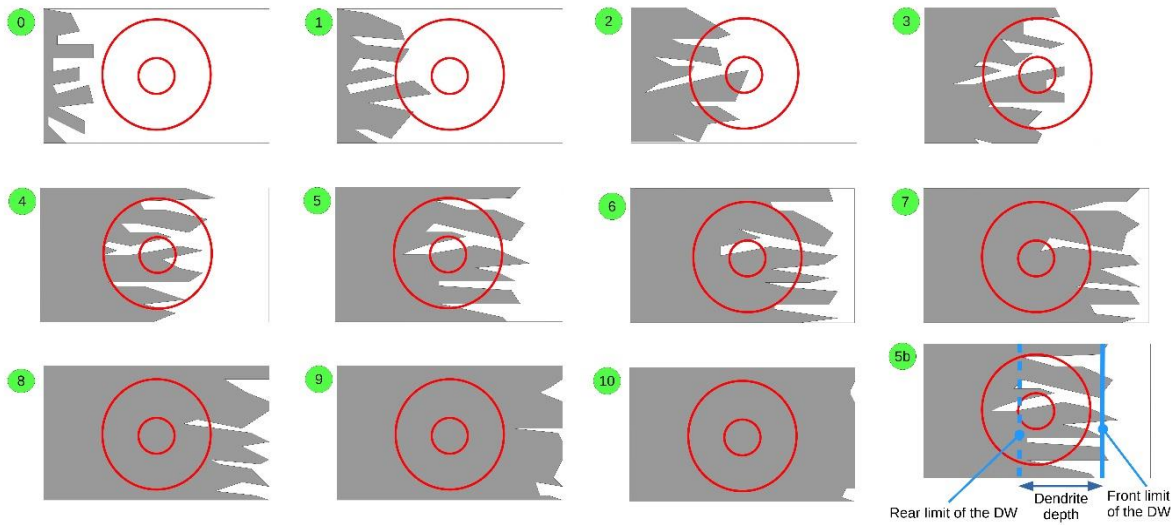

Figure 5. qualitative pictures of the successive positions of the domain wall in the case of a dendritic shape. Between two successive pictures, the mean movement is the same.

The ratio between the two velocities can be roughly described the following way: under the laser spot, the reversal starts when the front line of the domain wall arrives on the spot and finishes when the rear line of the domain wall reaches the other side of the spot. It means that the traveling distance of the front line is $2\omega_0+D$, where $\omega_0$ is the radius of the spot and D the depth of the dendrites (figure 5). So, the transit time in the spot is given by:

$$\tau = (2\omega_0+D)/v \qquad (3)$$

where v is the domain wall propagation velocity of the front part for the magnetic field applied. Assuming dendrites of depth D = 20µm, which is what we get looking at Kerr pictures (see figure 7), we have a good agreement with this basic model.

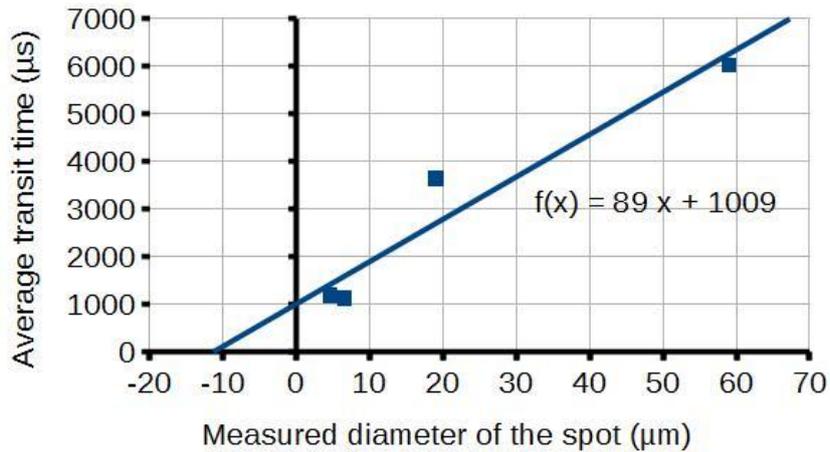

Figure 6. Average transit time obtained for an applied field of 1.95 mT. According to the linear fit, velocity would be 11 mm.s$^{-1}$, in good agreement with the velocity deduced from Kerr pictures for this field. The depth of the dendrite would be D = 11 µm.

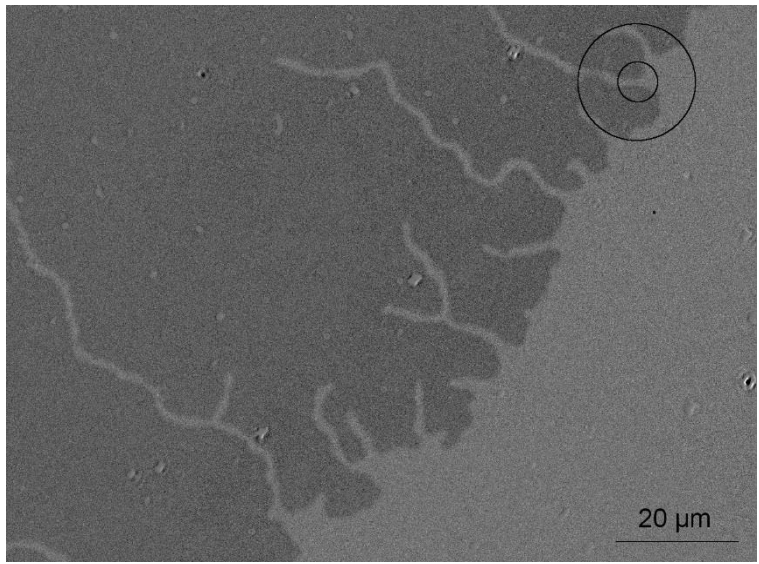

Figure 7. Kerr picture showing the shape of a domain wall after propagation induced by a 1.94 mT magnetic field. The circles in the upper part of the domain have a diameter of 6.5 µm and 19 µm, it gives an idea of the real figure going through the laser spot. The spot of diameter 58 µm has not been drawn as it covers most of the viewed area.

## IV. CONCLUSION

Looking at the transit time of the Kerr signal in a laser spot focused on a thin magnetic film is an interesting way to measure magnetic domain wall velocities. We have compared this method with the usual one using Kerr pictures: there is a shift between the curves obtained using the two ways. This shift disappears as the size of the spot is increased. We have explained this shift by the dendritic shape of the domain wall. As a result, we could get a very good agreement between both ways.

These results open a new way to check domain wall propagation in a magnetic thin film. One of the interesting features of this method is the real-time detection[30]. Indeed, through the usual way using Kerr pictures, one acquires a picture before and after a magnetic pulse, but there is no information about what happens during the pulse. Here, it becomes possible to get some information about the domain wall's characteristics during propagation.

**ACKNOWLEDGMENT**

The authors would like to thank the support from the program of China Scholarships Council (No. 201906020020), and the supports by the projects from National Natural Science Foundation of China (No. 12004024).

**REFERENCES**
[1]S. S. P. Parkin, M. Hayashi, and L. Thomas, Science **320,** 190 (2008).
[2]F. Riente, G. Turvani, M. Vacca, and M. G. Graziano, IEEE trans. on Emerging Topics in Computing **10,** 1216 (2022).
[3]J. H. Franken, H. J. M. Swagten, and B. Koopmans, Nature Nanotechnology **7,** 499 (2012).
[4]P. Diona, L. Gnoli, and F. Riente, IEEE Trans. on Electron Devices **69,** 3675 (2022).
[5]X. Zhang, N. Vernier, Z. Cao, Q. Leng, A. Cao, D. Ravelosona, and W. Zhao, Nanotechnology **29,** 365502 (2018).
[6]D. A. Allwood, G. Xiong, M. D. Cooke, C. C. Faulkner, D. Atkinson, N. Vernier, and R. P. Cowburn, Science **296,** 2003 (2002).
[7]D. A. Allwood, G. Xiong, C. C. Faulkner, D. Atkinson, D. Petit, and R. P. Cowburn, Science **309,** 1688 (2005).
[8]S. Breitkreutz, J. Kiermaier, I. Eichwald, C. Hildbrand, G. Csaba, D. Schmitt-Landsiedel, and M. Becherer, IEEE Trans. on Magnetics **49,** 4464 (2013).
[9]I. Eichwald, S. Breitkreutz, G. Ziemys, G. Csaba, W. Porod, and M. Becherer, Nanotechnology **25,** 335202 (2014).
[10]E. Varga, G. Csaba, G. H. Bernstein, and W. Porod, IEEE Trans. on Magnetics **50,** 2303804 (2014).
[11]J. Jaworowicz, N. Vernier, J. Ferré, A. Maziewski, D. Stanescu, D. Ravelosona, A. S. Jacqueline, C. Chappert, B. Rodmacq, and B. Diény, Nanotechnology **20,** 215401 (2009).
[12]W. H. Brigner, N. Hassan, L. Jiang-Wei, X. Hu, D. Saha, C. H. Bennett, M. J. Marinella, J. A. C. Incorvia, F. Garcia-Sanchez, and J. S. Friedman, IEEE Trans. on Electron Devices **66,** 4970 (2019).

[13]J. McCord, Journal of Physics D: Applied Physics **48,** 333001 (2015).

[14]W. Döring, **3,** 373 (1948).

[15]L. Thomas, R. Moriya, C. Rettner, and S. S. P. Parkin, Science **330,** 1810 (2010).

[16]F. Cayssol, D. Ravelosona, C. Chappert, J. Ferré, and J. P. Jamet, Physical Review Letters **92,** 107202 (2004).

[17]N. T. Hai, Z.-Y. Chen, I. Kindiak, R. C. Bhatt, L.-X. Ye, T.-h. Wu, K. A. Zvezdin, L. Horng, and J.-C. Wu, J. Mag. Magn. Mat. **546,** 168776 (2022).

[18]R. C. Bhatta, Y.-Y. Chenga, L.-X. Yea, N. Trong Haic, J.-C. Wuc, and T.-h. Wua, J. Mag. Mag. Mat. **513,** 167120 (2020).

[19]C. Burrowes, A. P. Mihai, D. Ravelosona, J. V. Kim, C. Chappert, L. Vila, A. Marty, Y. Samson, F. Garcia-Sanchez, L. D. Buda-Prejbeanu, I. Tudosa, E. E. Fullerton, and J. P. Attané, Nature Physics **6,** 17 (2010).

[20]C. Burrowes, N. Vernier, J. P. Adam, L. Herrera Diez, K. Garcia, I. Barisic, G. Agnus, S. Eimer, J.-V. Kim, T. Devolder, A. Lamperti, R. Mantovan, B. Ockert, E. E. Fullerton, and D. Ravelosona, Applied Physics Letters **103,** 182401 (2013).

[21]J. Grollier, P. Boulenc, V. Cros, A. Hamzić, A. Vaurès, A. Fert, and G. Faini, Applied Physics Letters **83,** 509 (2003).

[22]N. Vernier, D. A. Allwood, D. Atkinson, M. D. Cooke, and R. P. Cowburn, Europhysics Letters (EPL) **65,** 526 (2004).

[23]J. P. Adam, N. Vernier, J. Ferré, A. Thiaville, V. Jeudy, A. Lemaître, L. Thevenard, and G. Faini, Physical Review B **80,** 193204 (2009).

[24]P. Möhrke, T. A. Moore, M. Kläui, J. Boneberg, D. Backes, S. Krzyk, L. J. Heyderman, P. Leiderer, and U. Rüdiger, Journal of Physics D: Applied Physics **41,** 164009 (2008).

[25]T. A. Moore, P. Möhrke, L. Heyne, A. Kaldun, M. Kläui, D. Backes, J. Rhensius, L. J. Heyderman, J.-U. Thiele, G. Woltersdorf, A. Fraile Rodríguez, F. Nolting, T. O. Menteş, M. Á. Niño, A. Locatelli, A. Potenza, H. Marchetto, S. Cavill, and S. S. Dhesi, Physical Review B **82,** 094445 (2010).

[26]G. S. D. Beach, C. Nistor, C. Knutson, M. Tsoi, and J. L. Erskine, Nature Materials **4,** 741 (2005).

[27]S. Lemerle, J. Ferré, C. Chappert, V. Mathet, T. Giamarchi, and P. Le Doussal, Physical Review Letters **80,** 849 (1998).

[28]P. Chauve, T. Giamarchi, and P. Le Doussal, Physical Review B **62,** 6241 (2000).

[29]R. Diaz Pardo, W. Savero Torres, A. B. Kolton, S. Bustingorry, and V. Jeudy, Physical Review B **95,** 184434 (2017).

[30]T. Xing, N. Vernier, X. Y. Zhang, Y. G. Zhang, and W. S. Zhao, (to be published).